\documentclass{article}

\usepackage{arxiv}
\usepackage{amssymb}
\usepackage[utf8]{inputenc} 
\usepackage[T1]{fontenc}    
\usepackage{hyperref}       
\usepackage{url}            
\usepackage{booktabs}       
\usepackage{amsfonts}       
\usepackage{nicefrac}       
\usepackage{microtype}      
\usepackage{lipsum}		
\usepackage{graphicx}
\usepackage{natbib}
\usepackage{doi}
\usepackage{amsmath}

\title{Overlap and topology shape synergy in collective knowledge integration}

\author{Daeseong Kim\\
	Graduate School of Culture and Information Science,
    Doshisha University\\
    Kyotanabe, Kyoto, Japan\\
    \texttt{ctmp0014@mail4.doshisha.ac.jp} \\
	\And
	\href{https://orcid.org/0000-0002-0468-1179}{\includegraphics[scale=0.06]{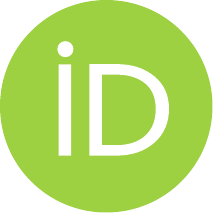}\hspace{1mm}Masato S.~Abe}\\
	Graduate School of Culture and Information Science, Doshisha University\\
Kyotanabe, Kyoto, Japan\\
Center for Advanced Intelligence Project, RIKEN, Tokyo, Japan\\
CBS-TOYOTA Collaboration Center, RIKEN, Wako, Saitama, Japan\\
	\texttt{maabe@mail.doshisha.ac.jp} \\
}

\renewcommand{\shorttitle}{Overlap and Topology Shape Synergy in Collective Knowledge Integration}

\hypersetup{
pdftitle={Overlap and topology shape synergy in collective knowledge integration},
pdfauthor={Daeseong Kim and Masato S.~Abe},
pdfkeywords={knowledge diversity, network science, small-world networks, adjacent possible, exploration}
}

\begin{document}
\maketitle

\begin{abstract}
	Collective discovery often requires the integration of partially overlapping knowledge, yet the beneficial level of overlap and the role of background topology remain unclear.
    We model individual knowledge as connected subnetworks of a common knowledge space and distinguish internal synergy, in which integration shortens paths between concepts, from external synergy, in which an unknown concept connects to the shared and all individual-specific regions.
    Internal synergy peaked at intermediate overlap across random, scale-free, modular, and small-world backgrounds, reflecting a balance between complementarity and common ground.
    External synergy was strongly topology dependent and was most pronounced when small-world structure maintained overlapping local boundaries.
    Rewiring separated the total supply of synergistic nodes from their concentration at the accessible boundary; moderate rewiring could increase structural opportunities while dispersing them among more external alternatives and reducing their discovery.
    Group-size effects on internal synergy depended on topology, whereas strict higher-order external synergy declined as its all-region criterion became more restrictive.
    Collective discovery therefore depends jointly on knowledge overlap and on how complementary knowledge regions are organized within the surrounding knowledge space.
\end{abstract}


\section{Introduction}
New knowledge is rarely created as an isolated element.
Scientific ideas, technologies, and solutions more often emerge through the recombination of previously established concepts and capabilities \citep{Weitzman1998,fleming2001,Arthur2009,youn2015}.
As the stock of available knowledge has expanded and expertise has become increasingly specialized, discovery has also become progressively collective \citep{jones2009}.
Teams now play a central role in producing both cumulative advances and disruptive contributions \citep{cowan2003, Wuchty2007, Rzhetsky2015, Hall2018, wu2019}, and their composition can shape which ideas are generated and which become successful \citep{Roger2005}.
The resulting question is therefore not simply whether groups outperform individuals, but under what conditions integration reorganizes relations within existing knowledge and alters how collective knowledge is positioned relative to potential discoveries \citep{woolley2010, muthukrishna2016, Riedl2021}.

Diversity is often regarded as a source of such collective advantage because individuals with different knowledge, experiences, and problem-solving strategies can contribute complementary resources \citep{Jehn1999, hong2004, dahlin2005, Page2007}.
Empirical studies likewise associate diverse expertise and network positions with greater originality, creativity, or team productivity \citep{reagans2001,fleming2007,van2023}.
Yet diversity alone does not guarantee successful integration.
Highly distinct knowledge can hinder mutual understanding and coordination, whereas highly similar knowledge provides common ground but leaves less complementary knowledge to recombine \citep{van2004, dahlin2005, cronin2007}.
Effective integration may therefore require a balance between commonality and distinctiveness.
From the standpoint of discovery, collective advantage consequently depends not only on how knowledge is distributed across individuals, but also on how existing knowledge is relationally organized and how that organization positions the collective with respect to concepts beyond its current knowledge boundary.

Because these potential advantages depend on relations among knowledge elements rather than merely on their presence or absence, understanding them requires a representation that preserves such relational structure.
Networks provide a natural representation for this purpose.
In a knowledge or semantic network, nodes represent concepts and edges represent meaningful relations, allowing individual knowledge to be described not only by which concepts are possessed but also by how they are connected \citep{kenett2014,abbott2015, kenett2016, benedek2017}.
Integrating partially distinct individual networks can then have two structural consequences: it can reorganize relations among already known concepts by creating shorter routes, and it can change how unknown concepts are related to shared and individual-specific regions of collective knowledge.
The latter is related to Koestler's  notion of \textit{bisociation}, in which creative acts bring previously distinct frames of thought into a new relation \citep{koestler1964}, and to the adjacent possible, in which the current organization of knowledge constrains which novelties become reachable next \citep{Kauffman2000,Iacopini2018, di2025}.
Collective integration may therefore matter not only because it increases the amount of known material, but also it reorganizes both existing relations and the boundary of potential novelty.

Recent studies have shown that network integration can generate structural synergy by creating paths or functional advantages unavailable in the constituent networks alone, and that the magnitude of these benefits depends on network topology \citep{ezaki2024, luppi2024}.
However, it remains unclear how the degree of overlap between partially shared knowledge structures shapes the structural advantages that emerge from their integration.
Overlap simultaneously determines how much knowledge individuals share and how much remains unique to each individual, suggesting that the consequences of integration cannot be understood from diversity or similarity alone.
Moreover, these consequences may appear in different parts of the knowledge structure: integration may reorganize relations among already known concepts, while also changing how the collective knowledge structure is related to concepts outside its current boundary.
Our aim is therefore to determine how overlap and background topology jointly shape the structural consequences of knowledge integration, both within the existing knowledge structure and at its boundary.

To address this problem, we develop a network-based model in which
individuals possess partially overlapping subnetworks of a common
knowledge space.
We distinguish \textit{internal synergy}, arising from
integration-induced path shortening, from \textit{external synergy},
arising when an unknown concept spans shared and individual-specific
knowledge regions.
This framework allows us to examine how knowledge overlap and background
topology jointly shape structural advantages within collective knowledge
and at its boundary.

\section{Results}
\subsection{Network model and synergy definitions}
\begin{figure*}
    \centering
    \includegraphics[width = \textwidth]{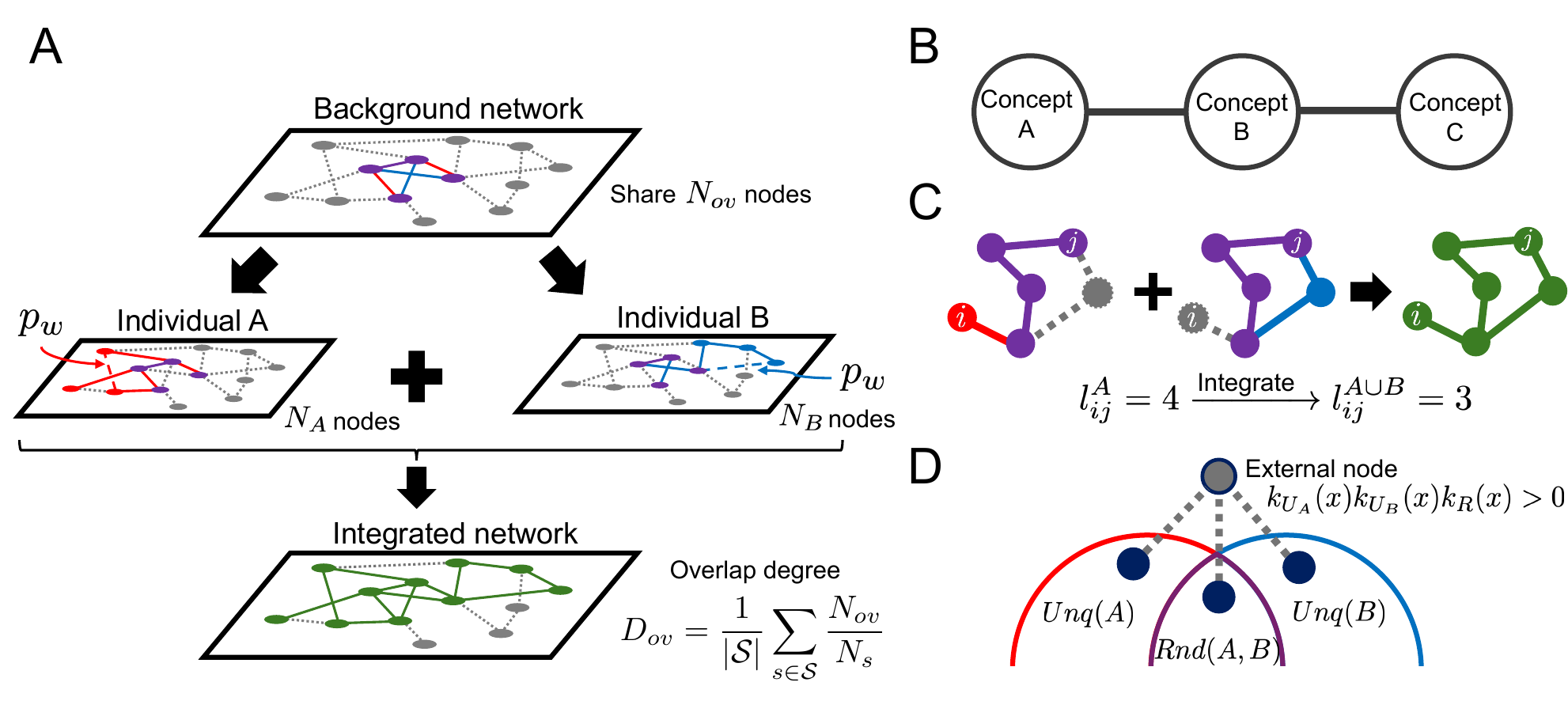}
    \caption{Schematic illustration of the network generation process and definition of synergies.
    (A) Let $\mathcal{S}=\{A,B\}$ denote the set of individuals.
    Each individual $s\in\mathcal{S}$ possesses a connected subnetwork of a common background network, with the subnetworks of A and B shown in red and blue, respectively, and their shared region in purple.
    The networks share $N_{ov}$ nodes, while $p_w$ controls additional within-individual wiring; shared nodes are identical across individuals although their relations may differ.
    Their union forms the integrated network.
    Gray dashed edges denote background relations not contained in the corresponding individual network.
    (B) Nodes represent knowledge elements and edges represent relations.
    (C) Internal synergy is an integration-induced reduction in shortest-path distance.
    (D) An external node is structurally synergistic when it connects to both individual-specific regions and the shared region.}
    \label{fig:fig1}
\end{figure*}
We represented potentially accessible knowledge as a background network whose nodes denote concepts and whose edges denote relations (Fig.~\ref{fig:fig1}A and B).
Each individual possessed a connected subnetwork of $N_s$ nodes, representing the set of concepts available to that individual.
A shared set of $N_{ov}$ nodes represented common knowledge, and the remaining nodes were individual-specific.
For equal-sized individuals, overlap degree was $D_{ov} = N_{ov}/N_s$; increasing $D_{ov}$ therefore increased common knowledge while reducing complementarity.

Knowledge, however, differs not only in which concepts are possessed but also in how richly those concepts are relationally organized.
After constructing each connected node set, we therefore retained additional background edges among selected nodes with probability $p_w$.
This parameter controls the relational richness of individual knowledge: low $p_w$ produces relatively sparse relational structures, whereas high $p_w$ represents individuals who possess a larger fraction of the relations that potentially connect their known concepts.
The collective network was formed as the union of the concepts and relations possessed by individuals.

We first quantified internal synergy using integration-induced shortest-path reduction, following \citet{luppi2024}.
In a knowledge network, path length reflects the number of relational steps separating concepts, so a shorter path indicates greater structural proximity and potentially easier access between them \citep{abbott2015,benedek2017}.
A node pair was therefore classified as internally synergistic when integration produced a path shorter than that available in either individual network (Fig.~\ref{fig:fig1}C).
Internal synergy thus measures how relations distributed across individuals reorganize existing knowledge into more direct collective connections.

We next defined external synergy to characterize the structural position of potential discoveries outside the collective network.
For two individuals, the integrated network was partitioned into an A-specific region $U_A$ 
, a B-specific region $U_B$, and their shared region $R$.
An external node was classified as structurally synergistic when it had at least one edge to each of these three regions (Fig.~\ref{fig:fig1}D).
This criterion captures an unknown concept whose relational context simultaneously involves shared and complementary knowledge, consistent with a combinatorial view of novelty and the adjacent possible \citep{mednick1962, koestler1964, youn2015, Kauffman2000, Iacopini2018, di2025}.

Importantly, a structurally synergistic external node need not be inaccessible to either individual in isolation.
Rather, the criterion identifies an unknown concept whose relational context spans the common knowledge required for coordination and the distinct knowledge contributed by both individuals.
External synergy therefore measures the presence of integrative opportunities at the boundary of collective knowledge, rather than discoveries that become possible through collaboration.
Direct enumeration measured the supply of such nodes, whereas first-exit random walks measured how often they were discovered (see \hyperref[sec:matmethods]{Materials and Methods} for more details). 

\subsection{Internal synergy peaks at intermediate overlap}
\begin{figure*}
    \centering
    \includegraphics[width =\textwidth]{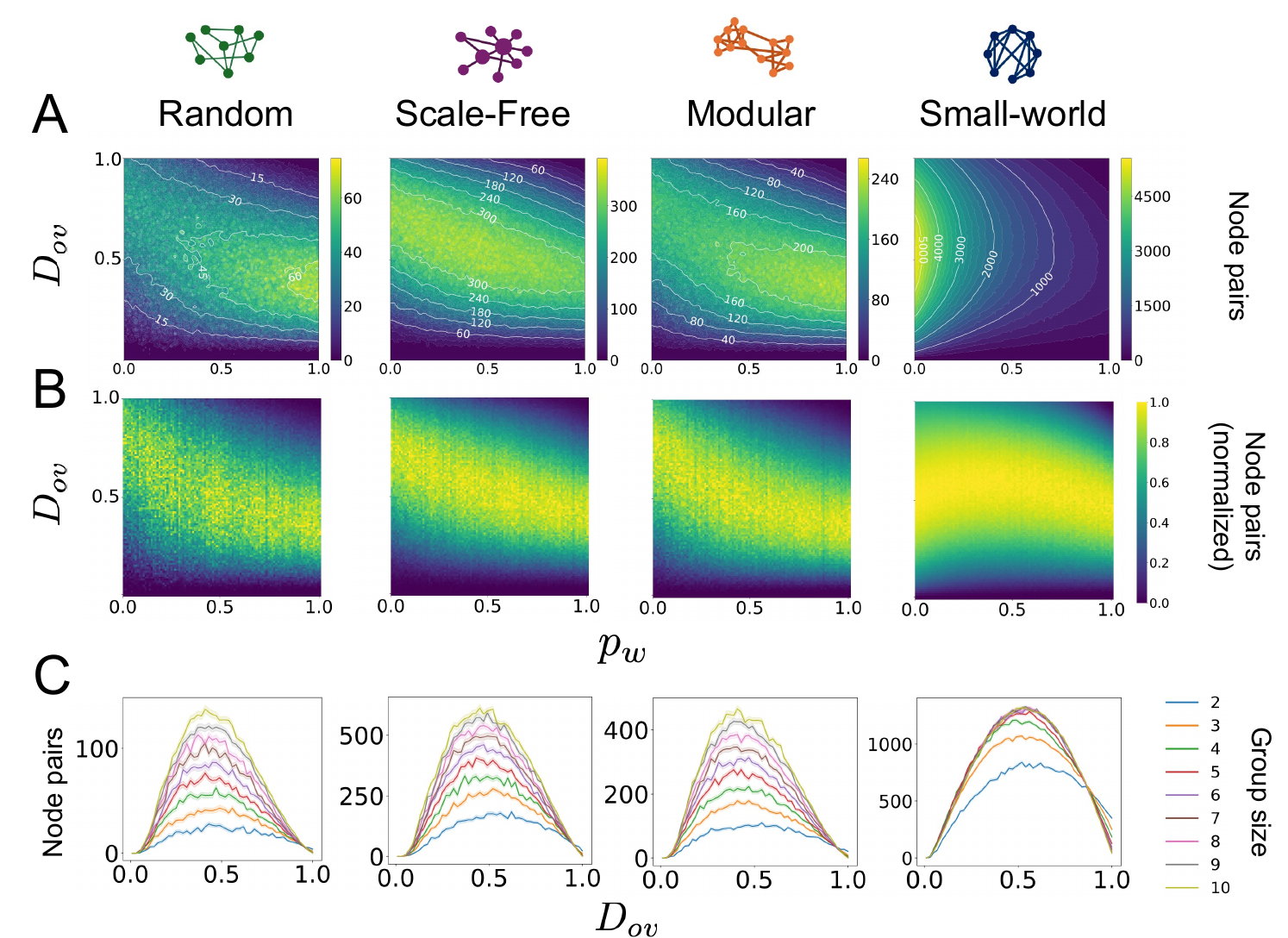}
    \caption{Emergence of internal synergy through collective integration of individual knowledge networks.
    (A) Number of synergistic node pairs as a function of $p_w$ and $D_{ov}$ in four background networks. Color scales differ among topologies and represent absolute counts.
    (B) Values normalized by the maximum across $D_{ov}$ for each $p_w$, highlighting the intermediate overlap region in which integration-induced shorter paths are most abundant.
    Each heatmap cell in (A) and (B) represents the mean across 100 independent network realizations.
    Sampling uncertainty was quantified using the standard error of the mean (SEM); absolute SEM, relative SEM, and representative mean $\pm$ SEM cross sections are reported in SI Appendix, Figs.~\ref{s:s1}-\ref{s:s3}.
    (C) Number of synergistic node pairs per individual as group size increases at $p_w = 0.5$; shading denotes SEM.}
    \label{fig:fig2}
\end{figure*}
Across random, scale-free, modular, and small-world backgrounds, internal synergy was weak near both extremes of overlap and largest at intermediate $D_{ov}$ (Fig.~\ref{fig:fig2}A).
At low overlap, the individual networks supplied many distinct nodes and paths but insufficient shared structure for those components to form numerous shortcuts.
At high overlap, integration was straightforward but the two networks contained too little non-redundant structure to create new routes.
The peak therefore marked a transition between a complementarity-dominated regime and a redundancy-dominated regime.

The absolute magnitude depended on both $p_w$ and background topology.
Small-world networks produced especially large numbers of synergistic pairs because spatially coherent subnetworks could be connected through shared nodes to create many shorter cross-region routes.
Nevertheless, normalization within each $p_w$ condition revealed a robust intermediate overlap optimum in all four topologies (Fig.~\ref{fig:fig2}B).
Supporting analyses of the same shortest-path classification were consistent with this trade-off. Uniquely inherited node pairs, for which the integrated shortest path was inherited from only one individual network, decreased with overlap, whereas redundant node pairs, for which both individual networks provided the same shortest-path distance, increased (SI Appendix, Figs.~\ref{s:s4} and \ref{s:s5}).

The same non-monotonic pattern remained when five individual networks were integrated (SI Appendix, Fig.~\ref{s:s6}).
Increasing group size increased the number of synergistic pairs per individual across all background topologies, but the dependence on group size differed across network structures (Fig.~\ref{fig:fig2}C).
In particular, small-world networks showed a saturation of per-individual synergy, whereas the other topologies continued to increase over the examined range.
The marginal effect of increasing group size therefore depended on background topology.
Despite these differences in magnitude, the non-monotonic dependence on overlap persisted across group sizes, with internal synergy remaining highest at intermediate $D_{ov}$.

\subsection{External synergy depends on local boundary organization}
\begin{figure*}
    \centering
    \includegraphics[width = \textwidth]{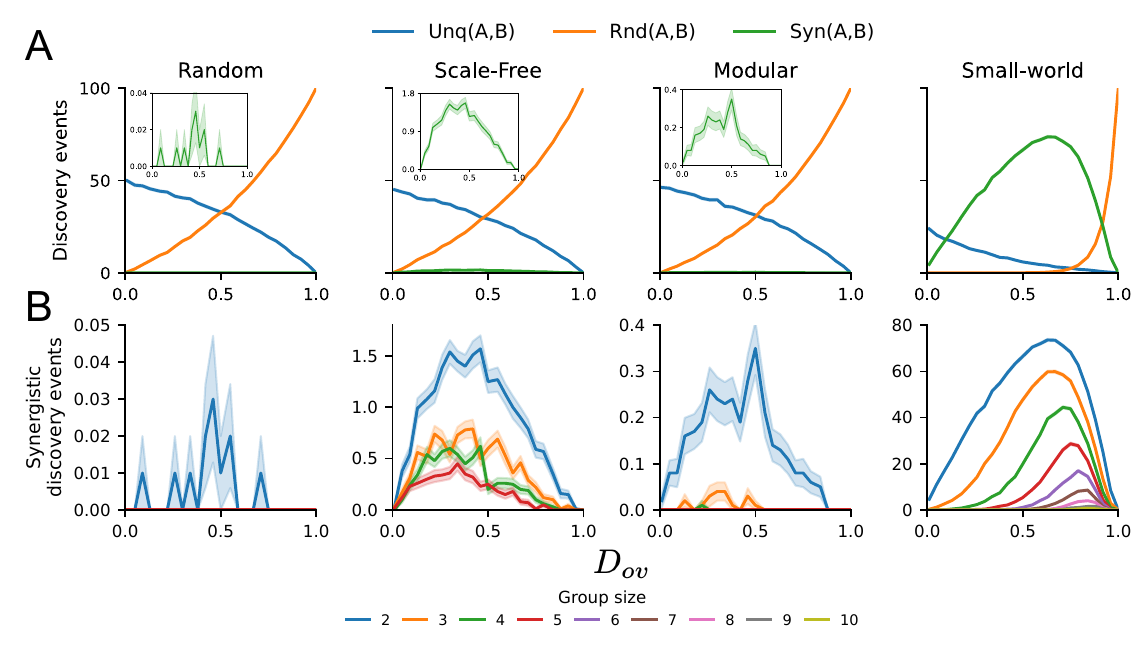}
    \caption{Emergence of external synergy through collective integration of individual knowledge networks.
    (A) First-exit discovery events classified as unique, redundant, or synergistic in random, scale-free, modular, and small-world backgrounds.
    The full classification, including mixed combinations of unique and redundant contributions, is shown in SI Appendix, Fig.~\ref{s:s7}.
    (B) Strict synergistic discovery events for increasing group size, evaluated up to 5 individuals for random, scale-free, and modular networks and up to 10 individuals for small-world networks.
    All panels use $p_w = 0.5$.
    The random, scale-free, and modular network panels use much smaller vertical scales than the small-world panel.
    Shading denotes SEM.}
    \label{fig:fig3}
\end{figure*}
The background topology produced a qualitative difference in discovery beyond current knowledge.
Because $p_w$ modifies internal wiring among already selected knowledge nodes without directly changing their background connections to external nodes, its effect on external synergy was expected to be limited.
Consistent with this expectation, external synergy patterns changed little across $p_w$ (SI Appendix, Fig.~\ref{s:s8}), and we therefore fixed $p_w = 0.5$ in the following analyses.

In the random network, first-exit discoveries shifted smoothly from individual-specific to shared region nodes as overlap increased, but synergistic events were nearly absent (Fig.~\ref{fig:fig3}A).
Scale-free and modular backgrounds showed similarly weak external synergy.
Although each knowledge region could be adjacent to many external nodes, their external neighborhoods rarely overlapped at the same node.
Consequently, few unknown nodes were simultaneously connected to the A-specific, B-specific, and shared regions.

Small-world networks instead displayed a pronounced inverse-U relationship.
At low overlap, the external neighborhoods of the two individual-specific regions overlapped only weakly because the shared region was small.
At high overlap, the individual-specific regions themselves became too small to maintain many such joint contacts.
Intermediate overlap retained a sufficiently large shared core while preserving distinct regions, allowing a common external node to lie at the joint boundary of all three components.
The result resembles a structural fold: partially overlapping regions remain distinguishable while creating an interface at which new combinations can emerge \citep{vedres2010, Uzzi2013, burt2022}.

The peak persisted as group size increased, but its magnitude fell sharply (Fig.~\ref{fig:fig3}B).
For $n$ individuals, the strict criterion required one external node to connect to the common region and to every individual-specific region, such that the order of the conjunction increased with group size.
Analysis of all possible combinations of connections to the different knowledge regions supported this interpretation (SI Appendix, Fig.~\ref{s:s9}).
At the overlap maximizing the strict criterion, the proportion of common-region discoveries connected to all individual-specific regions decreased from 90.3$\%$ for $n=2$ to 1.7$\%$ for $n=10$, whereas partial higher-order connections remained frequent.
Moreover, the fixed-order criterion requiring connections to at least two individual-specific regions remained broadly stable across group sizes.
This decline should therefore not be interpreted as a general reduction in creativity in larger teams, but as a consequence of applying an increasingly restrictive all-region conjunction.
Fixed-order or fractional-coverage criteria provide more comparable measures for groups of substantially different sizes.

\subsection{Boundary concentration predicts realized discovery}
\begin{figure*}
    \centering
    \includegraphics[width = \textwidth]{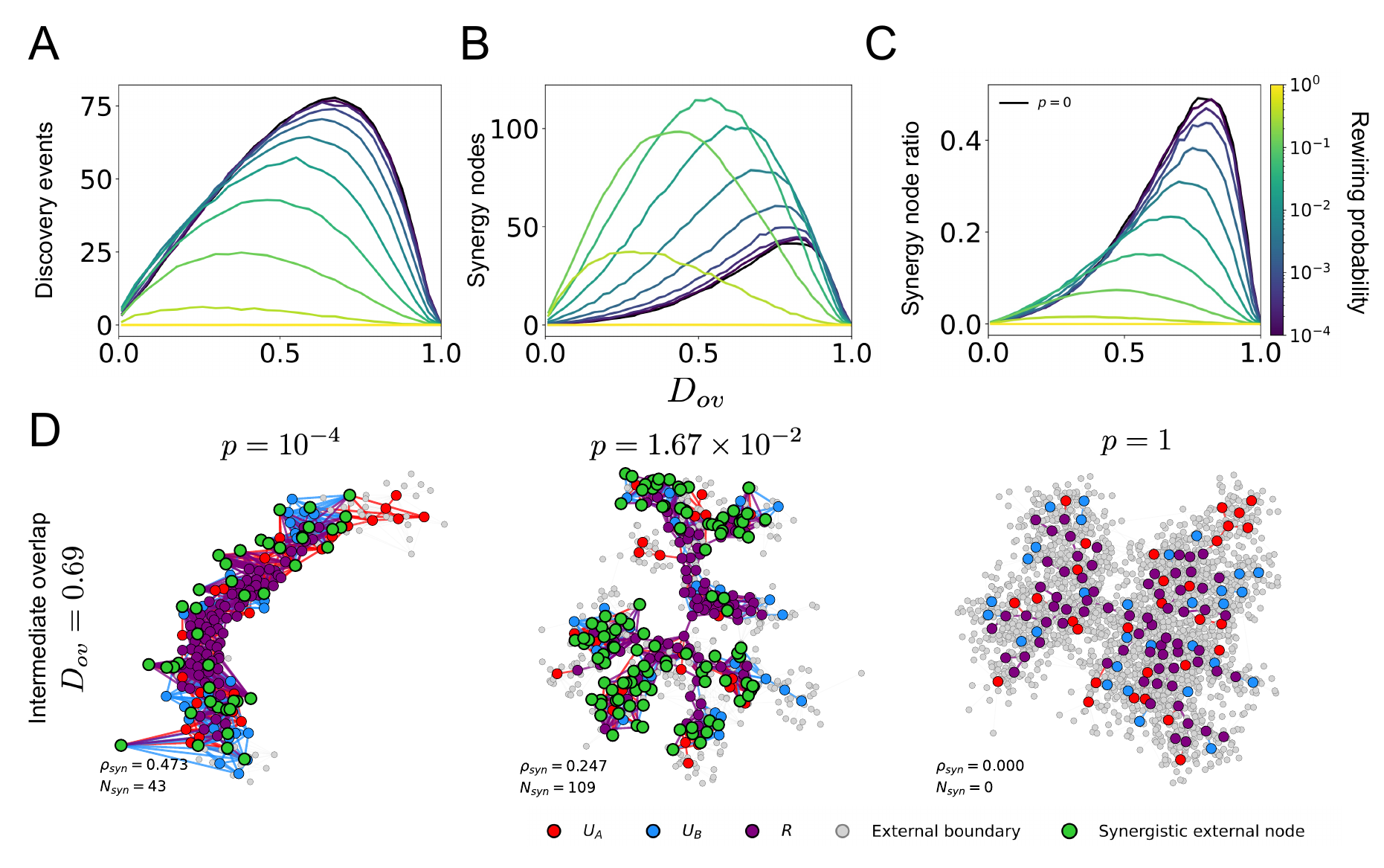}
    \caption{
    Boundary concentration and realization of external synergy during network rewiring.
    Watts-Strogatz rewiring probability $p$ was varied from a regular lattice toward increasingly random structure.
    (A) Synergistic first-exit discovery events as a function of overlap degree $D_{ov}$ for different $p$; the black line denotes $p=0$.
    (B) Absolute number of structurally synergistic external nodes, $N_{\textrm{syn}}$, irrespective of whether they were discovered.
    (C) Boundary concentration $\rho_{\textrm{syn}}$, the fraction of synergistic nodes among all external nodes adjacent to the integrated network.
    (D) Representative boundary configuration at fixed intermediate overlap for weak ($p = 10^{-4}$), moderate ($p=1.67\times 10^{-2}$), and strong ($p=1$) rewiring.
    Red, blue, and purple denote $U_A$, $U_B$, and $R$ respectively; gray denotes external boundary nodes and green structurally synergistic nodes.
    Labels report $\rho_{\mathrm{syn}}$ and $N_{\mathrm{syn}}$ for each realization.}
    \label{fig:fig4}
\end{figure*}

To identify the structural origin of the small-world effect, we varied the Watts-Strogatz rewiring probability $p$ from a regular lattice toward a random network (Fig.~\ref{fig:fig4}).
Realized synergistic discovery was strongest for regular and weakly rewired structures and declined as rewiring increased (Fig.~\ref{fig:fig4}A).
The absolute number of structurally synergistic external nodes behaved differently.
Moderate rewiring could increase the number of external nodes simultaneously connected to $U_A$, $U_B$, and $R$, and could shift the overlap at which their abundance was maximized (Fig.~\ref{fig:fig4}B).
A limited number of long-range links therefore created additional joint contacts between otherwise separated knowledge regions.

More opportunities, however, did not necessarily yield more discoveries.
We therefore measured the boundary concentration
\begin{equation}
    \rho_{\mathrm{syn}}
    =
    \frac{N_{\mathrm{syn}}}
    {N_{\mathrm{boundary}}},
\end{equation}
where $N_\mathrm{{boundary}}$ is the number of unknown external nodes adjacent to the integrated network and $N_{\mathrm{syn}}$ is the number of those nodes satisfying the structural synergy condition.
This quantity closely tracked the frequency of synergistic first-exit discoveries (Fig.~\ref{fig:fig4}C).

Representative realizations at fixed intermediate overlap make the distinction between absolute abundance and boundary concentration visible (Fig.~\ref{fig:fig4}D).
Under weak rewiring, shared and individual-specific knowledge remained locally coherent, so synergistic nodes occupied a relatively concentrated portion of the external boundary.
Moderate rewiring generated additional synergistic nodes, but it also expanded the surrounding external boundary, diluting those opportunities among a larger number of non-synergistic alternatives.
Under strong rewiring, local coherence was largely lost and joint contacts to all three knowledge regions became rare. 
The full set of representative configurations across both overlap and rewiring conditions is shown in SI Appendix, Fig.~\ref{s:s10}.

A phenomenological effective-contact approximation further showed that the overlap-dependent boundary-concentration curves across rewiring conditions can be summarized by the opposing effective exposure of individual-specific and shared regions (SI Appendix, Fig.~\ref{s:s11}).
The distinction parallels search in semantic networks, where random walk outcomes depend on the organization of locally available transitions \citep{abbott2015}, and small-world navigation, where global reachability can coexist with locally structured routes \citep{Watts1998, kleinberg2000}.
External synergy was therefore governed more directly by the local concentration of opportunities than by their total abundance.

\section{Discussion}
Our results show that the benefits of collective knowledge integration depend jointly on knowledge overlap and background topology.
Internal synergy was consistently maximized at intermediate overlap, whereas external synergy depended strongly on how shared and individual-specific regions were organized at the network boundary.
The number of structurally synergistic opportunities also did not necessarily predict how often they were encountered during exploration.

The intermediate-overlap optimum extends previous work showing that network integration can generate synergistic gains through shorter paths or improved function \citep{luppi2024,ezaki2024}.
It is also consistent with evidence that innovation benefits from a balance between relatedness and novelty \citep{cohen1990,nooteboom2007,Uzzi2013,yegros2015}.
In our model, shared knowledge therefore acts not only as redundancy but also as an interface through which complementary relational structures can be combined.

External synergy captures a different consequence of integration.
Combinatorial accounts of creativity and invention emphasize that novelty often emerges by bringing previously available elements into new relations \citep{mednick1962, simonton2003, youn2015}.
Our criterion identifies unknown concepts whose relational context simultaneously spans shared knowledge and knowledge specific to each collaborator.
Such concepts need not be inaccessible to either individual alone; rather, they mark positions where distributed knowledge becomes jointly relevant to a potential novelty.

The strong topology dependence of external synergy indicates that this joint relevance requires a particular organization of the knowledge boundary.
Small-world structure allowed the external neighborhoods of shared and individual-specific regions to overlap, whereas strong randomization dispersed these contacts across a broader boundary.
This mechanism parallels the idea that heterogeneous resources can be most productive when diversity is retained together with sufficient structural integration \citep{vedres2010,burt2022, fang2010}.
Our results extend this principle from networks of interacting agents to the topology of the knowledge being integrated.

The distinction between structural opportunity and realized discovery is equally important.
Moderate rewiring could increase the absolute number of synergistically positioned external nodes while reducing their concentration among all accessible boundary nodes.
First-exit discovery therefore tracked boundary concentration more closely than absolute abundance.
This result complements work showing that search outcomes depend not only on the availability of novel possibilities but also on how exploration is organized \citep{lazer2007,fang2010,foster2015}.

These results also suggest directions for connecting the model to empirical and emerging forms of knowledge production.
Large-scale semantic representations and individual semantic memory networks often exhibit locally clustered, short-path organization \citep{Steyvers2005,masucci2011,morais2013}, providing natural settings in which to test whether the joint-boundary mechanism occurs in real knowledge spaces.
The same structural question may become increasingly relevant in human-AI discovery, where AI systems expand access to information and generate candidate hypotheses across domains \citep{ding2025, zhang2025, xiao2026}.
Our results suggest that the benefit of such expanded access may depend not only on how much information becomes available, but also on how that information is organized relative to the knowledge already held by the human or collective system.

The present model is intentionally minimal.
Knowledge was represented as a static, undirected network, individual networks were generated through the same connected growth process, and exploration followed memoryless first-exit random walks.
The observed topology dependence may therefore reflect both background structure and the procedure by which individual knowledge is embedded within it.
Future work should test the proposed mechanisms in empirical semantic, scientific, and technological knowledge networks \citep{Steyvers2005,masucci2011, Uzzi2013, youn2015}.
More broadly, our results suggest that collective discovery depends not only on how much knowledge is shared or distributed, but on whether that knowledge is organized so that complementary regions can jointly relate to new possibilities.

\section{Materials and Methods}
\label{sec:matmethods}

\subsection{Networks and integration}
We generated background networks with $N= 10^5$ nodes and mean degree $\langle k \rangle \approx 20$ using NetworkX:
Erdős-Réyni random graphs with edge probability $20/(N-1)$ \citep{Erdos2006};
Barabási-Albert networks with $m=10$ \citep{Barabashi1999}; a 10-block stochastic block model with within- and between-block probabilities 0.00182 and 0.000022 \citep{holland1983}; and Watts-Strogatz networks with initial degree 20 and $p = 10^{-4}$ unless varied \citep{Watts1998}.

Let the background network be $G=(V,E)$.
Each individual $s$ was represented by a connected subgraph $G_s = (V_s,E_s)$ with $|V_s| = N_s$.
Individual networks shared a common node set
\begin{equation}
    R = \bigcap_{s\in\mathcal{S}}V_s,\qquad |R|=N_{ov}
\end{equation}
while the remaining nodes formed mutually exclusive individual-specific regions $U_s=V_s\setminus R$.
The shared and individual-specific regions were grown locally from background-network neighbors so that each individual network remained connected.
Edges used to maintain connectivity were selected independently for each individual, and additional background edges among selected nodes were retained independently with probability $p_w$.
Full network-generation procedures are described in SI Appendix.

The integrated knowledge network was the union of the individual networks,
\begin{equation}
    V_I = \bigcup_{s\in\mathcal{S}}V_s,\qquad E_I = \bigcup_{s\in\mathcal{S}}E_s.
\end{equation}
We quantified shared knowledge by
\begin{equation}
    D_{ov} = \frac{1}{|\mathcal{S}|}\sum_{s\in\mathcal{S}}\frac{N_{ov}}{N_s}.
\end{equation}

\subsection{Classification of internal integration effects}

For individual network $s$, let $\tilde{l}_{ij}^{s}$ denote the shortest-path distance between nodes $i$ and $j$, with $\tilde{l}_{ij}^s = +\infty $ when the pair is absent or disconnected.
Let $l_{ij}^I$ denote the corresponding distance in the integrated network.
A node pair was classified as internally synergistic when
\begin{equation}
    l_{ij}^I < \min_{s\in\mathcal{S}}\tilde{l}_{ij}^s.
\end{equation}
Thus, integration was synergistic when relations distributed across individual networks created a shorter path than was available in any constituent network.
Definitions of unique and redundant internal effects are provided in SI Appendix.

\subsection{Classification of external nodes}

For two individuals, the integrated network was partitioned into individual-specific regions $U_A$ and $U_B$ and the shared region $R$.
For an external node $x$, let
\begin{equation}
    k_S(x) = \sum_{j\in S}A_{xj}
\end{equation}
denote the number of background network edges connecting $x$ to region $S$.
We classified $x$ as structurally synergistic when 
\begin{equation}
    k_{U_A}(x)k_{U_B}(x)k_R(x)>0,
\end{equation}
that is, when it was connected to all three knowledge regions.

To quantify realized external synergy, we performed first-exit random walks on the integrated network.
A walk terminated when it first reached a node outside $V_I$, and the reached node was classified according to its connections to $U_A$, $U_B$, and $R$.
We separately measured the absolute number of structurally synergistic external nodes and their concentration among nodes adjacent to the integrated network.
Detailed external-node classifications are provided in SI Appendix.

\section{Acknowledgments}
We are grateful to Kimitaka Asatani for helpful discussions and valuable suggestions.








\bibliographystyle{unsrtnat}
\bibliography{references}  






\clearpage

\begin{center}
    {\LARGE\bfseries SI Appendix}
\end{center}
\vspace{1em}

\setcounter{section}{0}
\setcounter{subsection}{0}
\setcounter{figure}{0}
\setcounter{table}{0}
\setcounter{equation}{0}

\renewcommand{\thesection}{S\arabic{section}}
\renewcommand{\thesubsection}{\thesection.\arabic{subsection}}
\renewcommand{\thefigure}{S\arabic{figure}}
\renewcommand{\thetable}{S\arabic{table}}
\renewcommand{\theequation}{S\arabic{equation}}

\section{Supplementary Methods}
\subsection{Detailed generation of individual knowledge networks}
Let the background knowledge space be represented by an undirected graph
\begin{equation}
    G = (V,E),
\end{equation}
where nodes represent knowledge elements and edges represent potential relations between them.
For a group of $n$ individuals, individual $s$ was represented by a connected subgraph.
\begin{equation}
    G_s = (V_s,E_s),
\end{equation}
with target size $|V_s|=N_s$.

We first constructed a shared node set
\begin{equation}
    R=\bigcap_{s\in\mathcal{S}}V_s
\end{equation}
of prescribed size $|R| = N_{ov}$.
A seed node was selected uniformly at random from $V$ and added to $R$.
The shared region was then grown iteratively on the background network.
At each step, a node in 
\begin{equation}
    \Gamma(R)\setminus R
\end{equation}
was selected at random and added to $R$, where $\Gamma(S)$ denotes the set of background-network neighbors of nodes in $S$.
Growth continued until $|R| =N_{ov}$.

Although all individuals shared the same nodes set $R$, their edge sets within this region were constructed independently.
Whenever a new node was added during growth, each individual retained one background edge connecting that node to the previously acquired portion of its network.
Consequently, individuals could share the same knowledge elements while possessing different relational structures among them.

After constructing the shared region, each individual network was grown independently from $R$ until $|V_s| =N_s$.
For individual $s$, new nodes were sampled from the background boundary 
\begin{equation}
    \Gamma(V_s)\setminus\bigcup_{r=1}^{n}V_r,
\end{equation}
subject to the constraint that individual-specific nodes could not be assigned to more than one individual.
Each newly selected node was connected to the existing subnetwork through one background edge, ensuring that every $G_s$ remained connected.
This procedure produced mutually exclusive individual-specific regions
\begin{equation}
    U_s = V_s\setminus R,
\end{equation}
such that
\begin{equation}
    U_s\cap U_r = \varnothing\qquad (s\neq r).
\end{equation}

After the node sets were fixed, additional background edges between pairs of nodes belonging to the same individual network were retained independently with probability $p_w$.
Thus, for each individual,
\begin{equation}
    E_s \subseteq \left\{(i,j)\in E:\; i,j\in V_s \right\},
\end{equation}
with $p_w$ controlling the density of relations inherited from the background network.
Small $p_w$ therefore produced sparse connected subnetworks, whereas large $p_w$ retained a larger fraction of the background relations among the selected nodes.

The integrated knowledge network was defined as the union of the individual networks,
\begin{equation}
    G_I = (V_I,E_I),
\end{equation}
with
\begin{equation}
    V_I = \bigcup_{s=1}^{n}V_s,\qquad E_I = \bigcup_{s=1}^{n}E_s.
\end{equation}
Integration therefore introduced no relations that were absent from all individual networks; it combined only the knowledge elements and relations already possessed by at least one individual.

We quantified the amount of shared knowledge by the overlap degree
\begin{equation}
    D_{ov} = \frac{1}{|\mathcal{S}|}\sum_{s\in\mathcal{S}}\frac{N_{ov}}{N_s},
\end{equation}
where $|\mathcal{S}|$ is the number of individual networks being integrated.
Thus, $D_{ov}=0$ corresponds to vanishing common knowledge, whereas $D_{ov} = 1$ corresponds to completely overlapping node sets.
Accordingly, $D_{ov}$ controls the balance between shared and individual-specific knowledge in the collective network.

\subsection{Additional classification of internal integration effects}
For individual network $s$, let $V_s$ denote its node set and $l_{ij}^{s}$ denote the shortest-path distance between nodes $i$ and $j$.
To compare node pairs that may not be present in every individual network,
we define the extended shortest-path distance as
\begin{equation}
\tilde{l}_{ij}^{s}
:=
\begin{cases}
l_{ij}^{s},
& \{i,j\}\subseteq V_s
\text{ and } i \text{ and } j \text{ are connected},\\
+\infty,
& \text{otherwise}.
\end{cases}
\label{eq:extended_distance}
\end{equation}

Let $l_{ij}^I$ denote the shortest-path distance between the same
unordered node pair in the integrated network. We classified each
unordered pair $\{i,j\}\subseteq V_I$ according to the relationship
between its individual and integrated distances.

A node pair was classified as \textit{internally synergistic} when
integration generated a path shorter than that available in either
individual network:
\begin{equation}
l_{ij}^{I}
<
\min_{s\in\mathcal{S}}
\tilde{l}_{ij}^{s}.
\label{eq:internal_synergy}
\end{equation}

A node pair was classified as \textit{unique to individual A} when the
shortest path in the integrated network was inherited from individual A
and was shorter than the corresponding path available in individual B:
\begin{equation}
l_{ij}^{I}
=
\tilde{l}_{ij}^{A}
<
\tilde{l}_{ij}^{B}.
\label{eq:unique_A}
\end{equation}
Similarly, a node pair was classified as \textit{unique to individual B}
when
\begin{equation}
l_{ij}^{I}
=
\tilde{l}_{ij}^{B}
<
\tilde{l}_{ij}^{A}.
\label{eq:unique_B}
\end{equation}

A node pair was classified as \textit{redundant} when both individual
networks and the integrated network provided the same shortest-path
distance:
\begin{equation}
\tilde{l}_{ij}^{A}
=
\tilde{l}_{ij}^{B}
=
l_{ij}^{I}.
\label{eq:redundancy}
\end{equation}

\subsection{Additional classification of External synergy effects}
For two individual knowledge networks, the integrated network was
partitioned into three mutually distinct node regions:
the A-specific region
\begin{equation}
    U_A = V_A \setminus V_B,
\end{equation}
the B-specific region
\begin{equation}
    U_B = V_B \setminus V_A,
\end{equation}
and the shared region
\begin{equation}
    R = V_A \cap V_B.
\end{equation}
Nodes outside the integrated network,
$V_I=V_A\cup V_B$, were treated as unknown external nodes.

For an external node $x$, we defined the number of background-network
edges connecting it to region $S\in\{U_A,U_B,R\}$ as
\begin{equation}
    k_S(x)
    =
    \sum_{j\in S} A_{xj},
\end{equation}
where $A_{xj}$ is the adjacency matrix of the background network.
We then defined a binary indicator
\begin{equation}
    b_S(x)
    =
    \mathbf{1}[k_S(x)>0],
\end{equation}
which specifies whether $x$ is connected to at least one node in
region $S$.
Thus, each external node can be represented by the connection pattern
\begin{equation}
    \mathbf{b}(x)
    =
    \left(
    b_{U_A}(x),
    b_{U_B}(x),
    b_R(x)
    \right).
\end{equation}

An external node was classified as \textit{strictly synergistic} when
it was connected to all three regions,
\begin{equation}
    \mathbf{b}(x)=(1,1,1),
\end{equation}
or equivalently,
\begin{equation}
    k_{U_A}(x)k_{U_B}(x)k_R(x)>0.
\end{equation}
This criterion identifies external nodes whose relational context
simultaneously spans the knowledge specific to both individuals and
the knowledge shared between them.

For comparison with strict synergy, external nodes were further
classified according to the complete pattern of regional connections.
For two individuals, the nonempty connection patterns are
\begin{align}
    (1,0,0) &: U_A\text{-specific}, \nonumber\\
    (0,1,0) &: U_B\text{-specific}, \nonumber\\
    (0,0,1) &: R\text{-only}, \nonumber\\
    (1,1,0) &: U_A+U_B, \nonumber\\
    (1,0,1) &: U_A+R, \nonumber\\
    (0,1,1) &: U_B+R, \nonumber\\
    (1,1,1) &: \text{strict synergy}.
\end{align}
The first three classes correspond to external nodes associated with a
single structural region, whereas the intermediate classes represent
partial combinations of multiple regions.
Only the $(1,1,1)$ class satisfies the strict external-synergy
criterion.

\subsection{Group-size dependence of higher-order external synergy}
The strict external-synergy criterion becomes progressively more
restrictive as group size increases. For a group of $n$ individuals,
we represented the connection pattern of a discovered external node as
\begin{equation}
    \mathbf{b}
    =
    \left(b_1,b_2,\ldots,b_n,b_R\right),
\end{equation}
where $b_i=1$ indicates a connection to the individual-specific region
of individual $i$, and $b_R=1$ indicates a connection to the common
region. We defined the connection order
\begin{equation}
    r=\sum_{i=1}^{n} b_i
\end{equation}
as the number of individual-specific regions connected to the external
node. The strict criterion used in the main text therefore requires
$b_R=1$ and $r=n$: the external node must simultaneously connect to the
common region and to all $n$ individual-specific regions. For $n=2$,
$r=0$, $r=1$, and $r=2$ correspond respectively to the $Rnd$, $Unq$,
and $Syn$ classes used in the two-individual analysis.

The intermediate-overlap maximum of the strict criterion persisted from
$n=2$ to $n=10$, but its magnitude decreased sharply with group size
(Fig.~\ref{s:s9}A). This decline can arise for two conceptually different reasons:
external nodes connecting multiple knowledge regions may themselves
become rare, or such nodes may remain abundant while the requirement of
connecting to \emph{all} individual-specific regions becomes increasingly
difficult to satisfy. To distinguish these possibilities, we decomposed
discovery events according to their connection order at the value of
$D_{ov}$ that maximized the strict criterion for each group size.

Among discovery events connected to the common region, we distinguished
common-only events ($r=0$), partial events involving fewer than half of
the individual-specific regions
[$1\leq r<\lceil n/2\rceil$], partial events involving at least half but
not all individual-specific regions
[$\lceil n/2\rceil\leq r<n$], and strict events ($r=n$).
For $n=2$, the below-half partial class is empty, whereas the
at-least-half partial class consists of $r=1$ events and is therefore
equivalent to the $Unq$ category. The common-only class remains nonzero
but constitutes only a very small fraction at the strict-synergy peak.
As group size increased, the relative contribution of the strict class
declined, whereas partial connections involving a substantial fraction
of the individual-specific regions became increasingly dominant
(Fig.~\ref{s:s9}B).

To test whether the decline in strict external synergy was primarily a
consequence of this increasing conjunction order, we compared the strict
criterion with two relaxed criteria. The fixed-order criterion required
a connection to the common region and to at least two
individual-specific regions,
\begin{equation}
    b_R=1,\qquad r\geq 2,
\end{equation}
so that the required number of individual-specific regions remained
constant as group size increased. We additionally considered a
half-coverage criterion,
\begin{equation}
    b_R=1,\qquad
    r\geq \left\lceil 0.5n\right\rceil,
\end{equation}
for which the required connection order increased with group size but
less rapidly than under the strict criterion. For each criterion and
group size, we independently evaluated the maximum number of discovery
events over $D_{ov}$.

The maximum number of fixed-order events remained broadly stable as
group size increased, while the half-coverage criterion retained
substantial discovery counts even for the largest groups
(Fig.~\ref{s:s9}C). In contrast, strict events became increasingly rare because
the external node was required to connect simultaneously to every
individual-specific region. These results indicate that increasing group
size does not generally eliminate external nodes that combine multiple
knowledge regions. Rather, the sharp decline in strict external synergy
is largely attributable to the increasing order of the all-region
conjunction imposed by the strict definition.

\subsection{Visualization of boundary organization across overlap and rewiring}

To complement the quantitative analysis in Fig.~4, we visualized
representative realizations of the integrated knowledge network across
three levels of overlap and three Watts--Strogatz rewiring regimes
(Fig.~\ref{s:s10}).
The columns correspond to low, intermediate, and high overlap
($D_{ov}=0.20$, $0.69$, and $0.88$), whereas the rows correspond to
weakly rewired, moderately rewired, and strongly rewired background
networks ($p=10^{-4}$, $1.67\times10^{-2}$, and $1$, respectively).
The intermediate-overlap column ($D_{ov}=0.69$) is reproduced in
Fig.~\ref{fig:fig4}D to illustrate the effect of rewiring while holding knowledge
overlap fixed.

For each realization, we displayed the two individual-specific regions,
their shared region, and all unknown external nodes directly adjacent to
the integrated network.
An external node was classified as structurally synergistic when it had
at least one connection to each of the two individual-specific regions
and to the shared region.
The boundary concentration
$\rho_{\mathrm{syn}}$ denotes the fraction of these
structurally synergistic nodes among all unknown external nodes adjacent
to the integrated network, whereas $N_{\mathrm{syn}}$ denotes their
absolute number.
The snapshots are intended to illustrate local boundary organization
rather than to quantify differences among conditions; ensemble-level
changes in $N_{\mathrm{syn}}$, $\rho_{\mathrm{syn}}$, and
realized synergistic discovery are reported in Fig.~\ref{fig:fig4}.

At weak rewiring, shared and individual-specific knowledge regions remain
locally coherent, allowing synergistically positioned external nodes to
occupy a relatively concentrated portion of the accessible boundary.
Moderate rewiring can generate additional synergistic contacts and
increase $N_{\mathrm{syn}}$, while simultaneously expanding the external
boundary and thereby reducing their relative concentration
$\rho_{\mathrm{syn}}$.
Under strong rewiring, local coherence is largely lost and joint contacts
to all three knowledge regions become rare.

Variation across the columns additionally illustrates the role of
knowledge overlap.
At low overlap, the individual-specific regions are large but the shared
region is limited.
At high overlap, the shared region dominates while the
individual-specific regions become small.
Intermediate overlap preserves both shared and individual-specific
components, providing the structural conditions under which joint
external boundaries can form.

\subsection{Phenomenological effective-contact approximation for external synergy}

To obtain a compact representation of the overlap dependence of external
synergy, we considered a phenomenological effective-contact approximation
for the two-individual case. The purpose of this approximation is not to
derive the Watts--Strogatz result from network topology, but to determine
whether the non-monotonic dependence on overlap can be represented by a
simple competition between effective exposure to individual-specific and
shared knowledge regions.

For an external node with $k$ potential contacts, we denote the effective
per-contact probabilities of connection to the two individual-specific
regions by
\begin{equation}
    \theta_A(D_{ov};p)
    =
    \theta_B(D_{ov};p)
    =
    \theta_U(D_{ov};p),
\end{equation}
where
\begin{equation}
    \theta_U(D_{ov};p)
    =
    \phi_p
    \frac{1-D_{ov}}
    {2(1-D_{ov})+\eta_p D_{ov}^{2}},
\end{equation}
and the corresponding probability of connection to the shared region is
\begin{equation}
    \theta_R(D_{ov};p)
    =
    \phi_p
    \frac{\eta_p D_{ov}^{2}}
    {2(1-D_{ov})+\eta_p D_{ov}^{2}}.
\end{equation}
These quantities satisfy
\begin{equation}
    2\theta_U+\theta_R=\phi_p,
\end{equation}
so that $\phi_p$ controls the overall effective exposure to the three
knowledge regions, whereas $\eta_p$ controls how this exposure is
distributed between individual-specific and shared regions.

Let
\begin{equation}
    \theta_0=1-2\theta_U-\theta_R
\end{equation}
denote the probability that a contact does not terminate in any of the
three regions. Under independent contact trials, the probability that an
external node is connected to both individual-specific regions and to the
shared region is obtained by inclusion--exclusion:
\begin{align}
P_{\mathrm{Syn}}
={}&
1
-2(1-\theta_U)^k
-(1-\theta_R)^k
+(1-2\theta_U)^k
\nonumber\\
&+
2(1-\theta_U-\theta_R)^k
-\theta_0^k .
\end{align}

Because the quantity shown in Fig.~\ref{fig:fig4}C is the fraction of synergistic nodes
among external nodes already adjacent to the integrated network, we
condition the approximation on membership in this external boundary:
\begin{equation}
    P_{\mathrm{Syn}\mid\partial}
    =
    \frac{P_{\mathrm{Syn}}}
    {1-\theta_0^k}.
\end{equation}

For the weakly rewired condition ($p=10^{-4}$), increasing overlap shifts
effective exposure from the individual-specific regions toward the shared
region (Fig.~\ref{s:s11}A). Consequently, individual-specific-only events decrease
while shared-region-only events increase, whereas simultaneous exposure to
all three regions is maximized at intermediate overlap (Fig.~\ref{s:s11}B).

We next asked whether the same functional form could also summarize the
systematic change in boundary concentration with network rewiring. For
each rewiring probability $p$, the parameters $\phi_p$ and $\eta_p$ were
estimated by minimizing the squared difference between
$P_{\mathrm{Syn}\mid\partial}(D_{ov};p)$ and the simulated boundary
concentration $\rho_{\mathrm{syn}}(D_{ov};p)$.
Overlaying the fitted curves with the simulation results shows that the
two-parameter approximation reproduces both the reduction in peak
magnitude and the approximate shift of the peak toward lower overlap as
rewiring increases (Fig.~\ref{s:s11}C).

This agreement should be interpreted as a phenomenological reconstruction,
not as an independent prediction. Because $\phi_p$ and $\eta_p$ are fitted
separately to the boundary-concentration data for each rewiring condition,
the approximation demonstrates that the family of simulation curves can be
represented by a low-dimensional effective-contact description.
Furthermore, the independence assumption does not explicitly capture the
local correlations generated by small-world structure. The fitted
parameters therefore summarize these structural effects rather than
providing a first-principles derivation of them. At very large rewiring
probabilities, where synergistic boundary events are nearly absent, the
parameters also become weakly identifiable.

\section{Supplementary Figures}
\begin{figure}
\centering
\includegraphics[width=\textwidth]{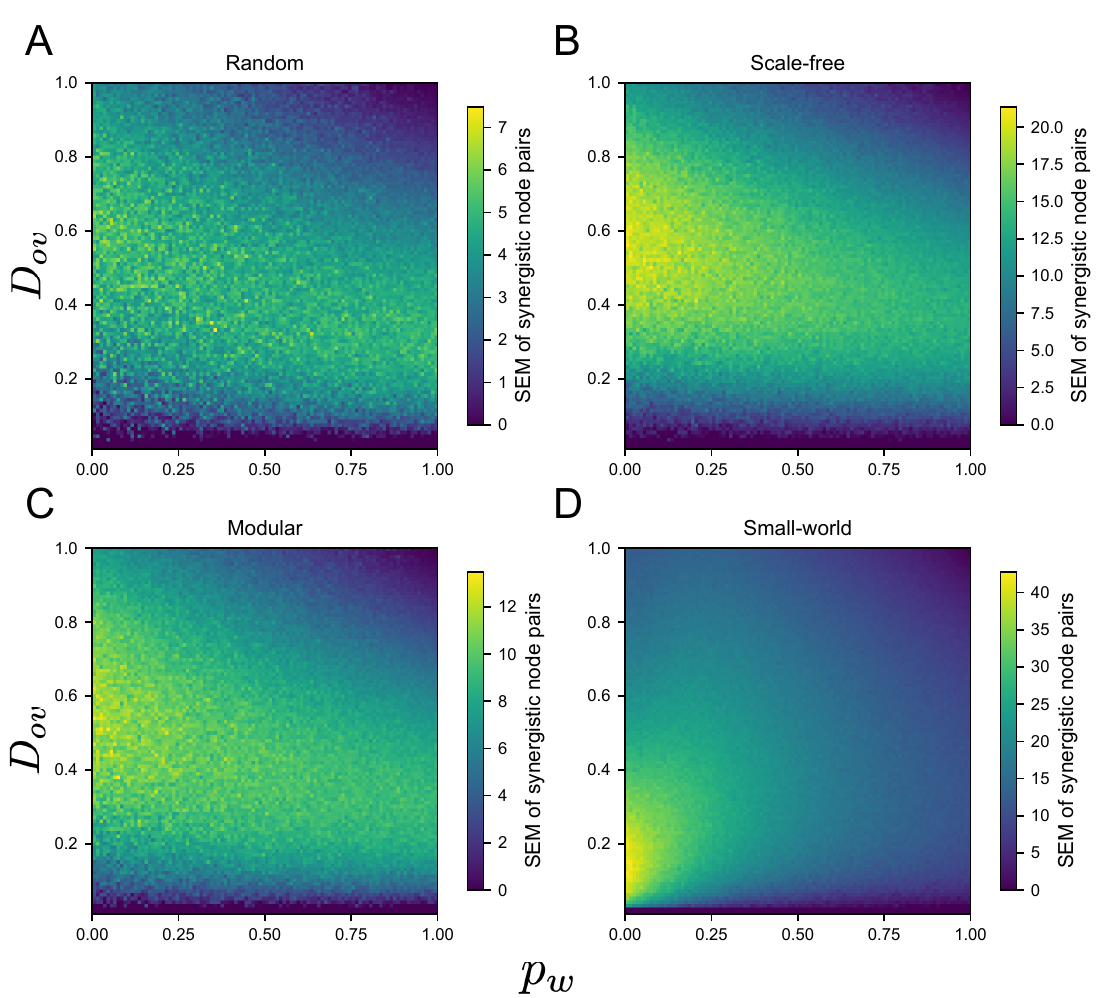}
    \caption{\textbf{Absolute sampling uncertainty of internal synergy.}
    Standard errors of the mean (SEM) for the number of synergistic node pairs across the parameter space examined in Fig.~\ref{fig:fig2}A.
    Each cell corresponds to one combination of overlap degree $D_{ov}$ and within-individual wiring probability $p_w$, and the SEM was calculated across 100 independently generated network realizations.
    Panels show random (A), scale-free (B), modular (C), and small world (D) background networks.}
    \label{s:s1}
\end{figure}

\begin{figure}
\centering
\includegraphics[width=\textwidth]{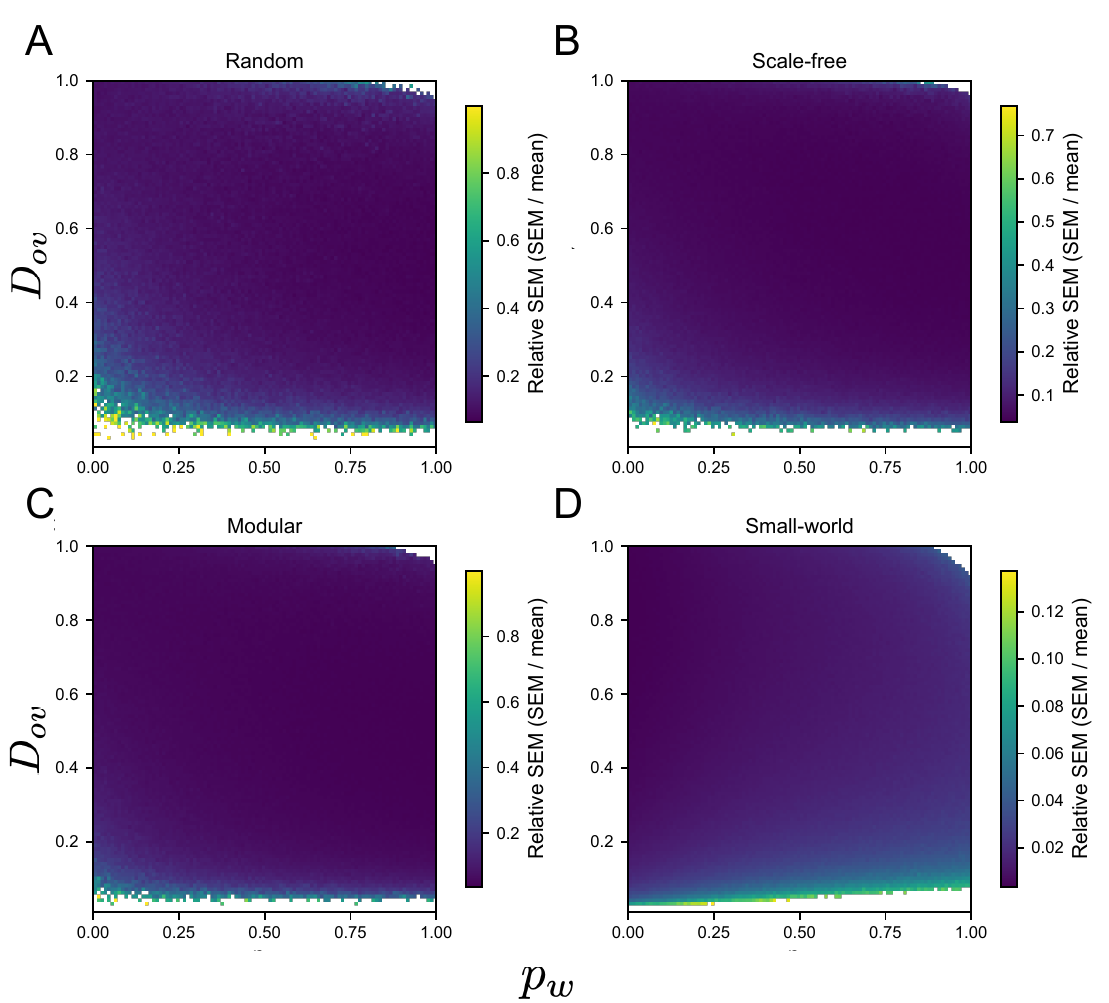}
    \caption{\textbf{Relative sampling uncertainty of internal synergy.}
    Relative sampling uncertainty was defined as
    $\mathrm{rSEM}(D_{ov},p_w)
    =
    \mathrm{SEM}(D_{ov},p_w)/
    \overline{S}(D_{ov},p_w)$,
    where $\overline{S}(D_{ov},p_w)$ is the mean number of synergistic node pairs
    and $\mathrm{SEM}(D_{ov},p_w)$ is its standard error across 100 independently
    generated network realizations.
    Thus, rSEM represents the sampling uncertainty relative to the magnitude
    of the corresponding mean.
    Cells for which the mean was below 1\% of the maximum mean within each
    background-network structure were omitted because the ratio becomes unstable
    as the denominator approaches zero.
    Panels show random (A), scale-free (B), modular (C), and small-world (D)
    background networks.}
    \label{s:s2}
\end{figure}

\begin{figure}
\centering
\includegraphics[width=\textwidth]{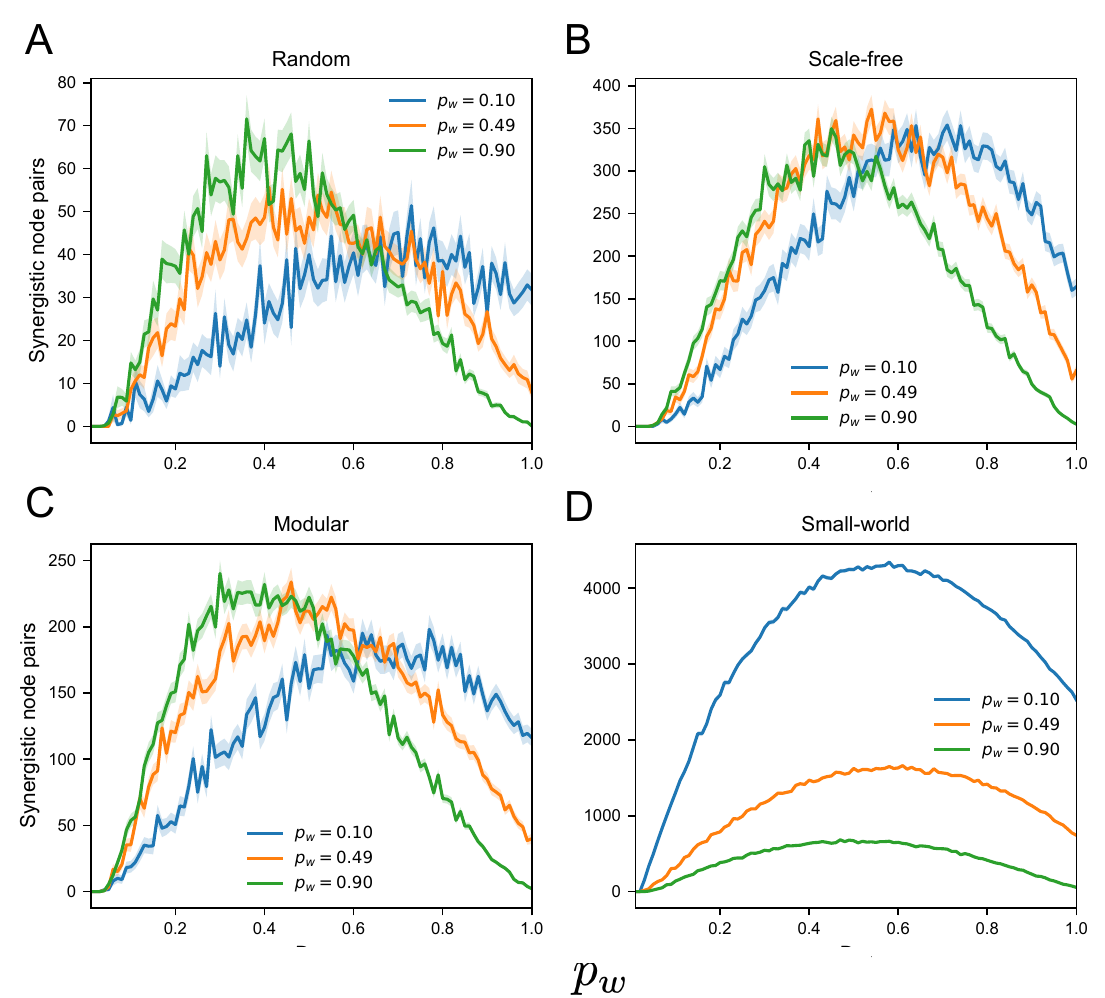}
    \caption{\textbf{Representative cross-sections of internal synergy with sampling uncertainty.}
    The mean number of synergistic node pairs is shown as a function of overlap degree $D_{ov}$ for representative within-individual wiring probabilities $p_w=0.1$, $0.5$, and $0.9$.
    Lines indicate means across 100 independently generated network realizations, and shaded regions indicate $\pm 1$ SEM.
    These cross-sections show that the intermediate-overlap maximum observed in Fig.~\ref{fig:fig2} is retained after accounting for sampling uncertainty.
    Panels show random (A), scale-free (B), modular (C), and small-world (D) background networks.}
    \label{s:s3}
\end{figure}

\begin{figure}
\centering
\includegraphics[width=\textwidth]{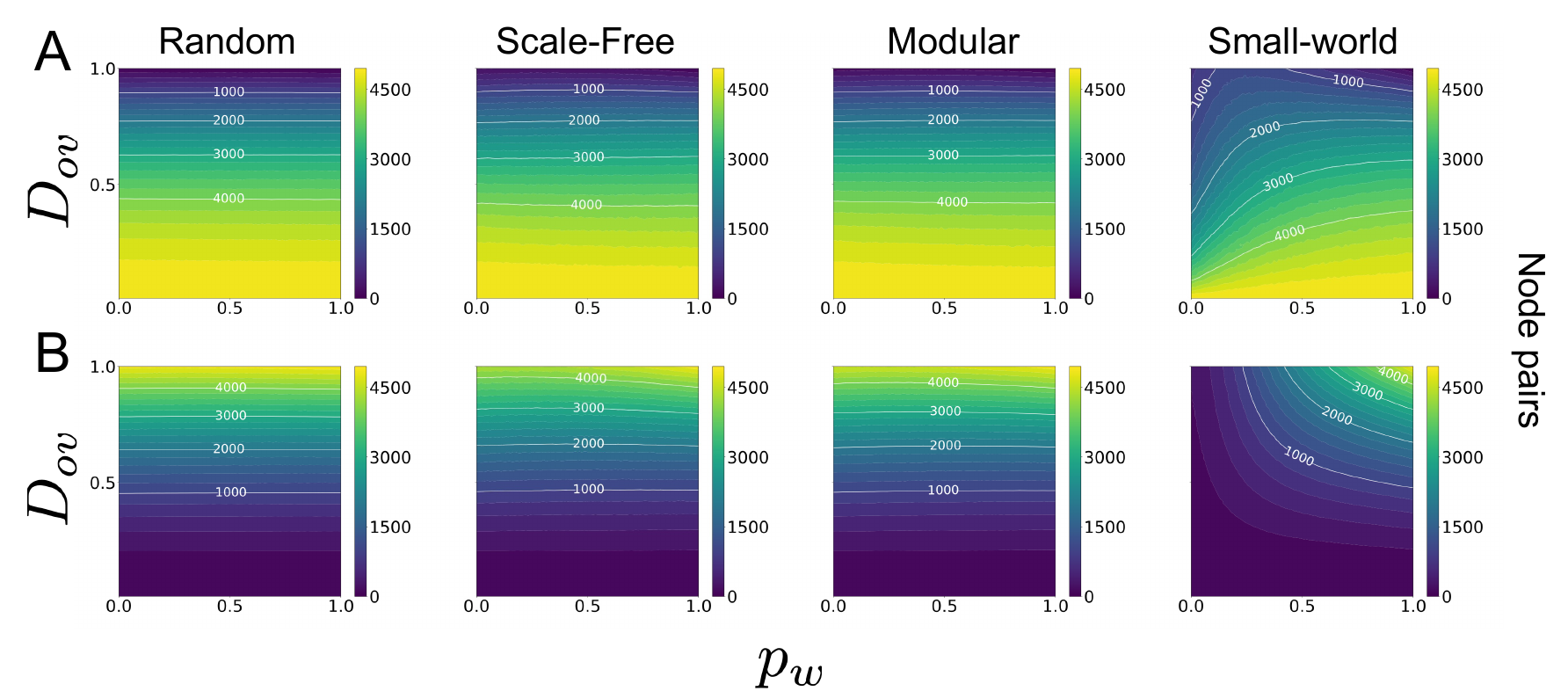}
    \caption{\textbf{Absolute numbers of unique and redundant node pairs after the integration of two individual knowledge networks.}
    Heatmaps show the mean number of node pairs classified as\textbf{(A)} unique and \textbf{(B)}redundant as functions of the within-individual wiring probability $p_w$ and the overlap degree $D_{ov}$. 
    Results are shown for random, scale-free, modular, and
    small world background networks. Unique and redundant classes were determined by comparing shortest-path lengths in the two individual networks with those in the integrated network, as defined in Materials and Methods. 
    Colors and white contours indicate the absolute number of classified node pairs.}
    \label{s:s4}
\end{figure}

\begin{figure}
\centering
\includegraphics[width=\textwidth]{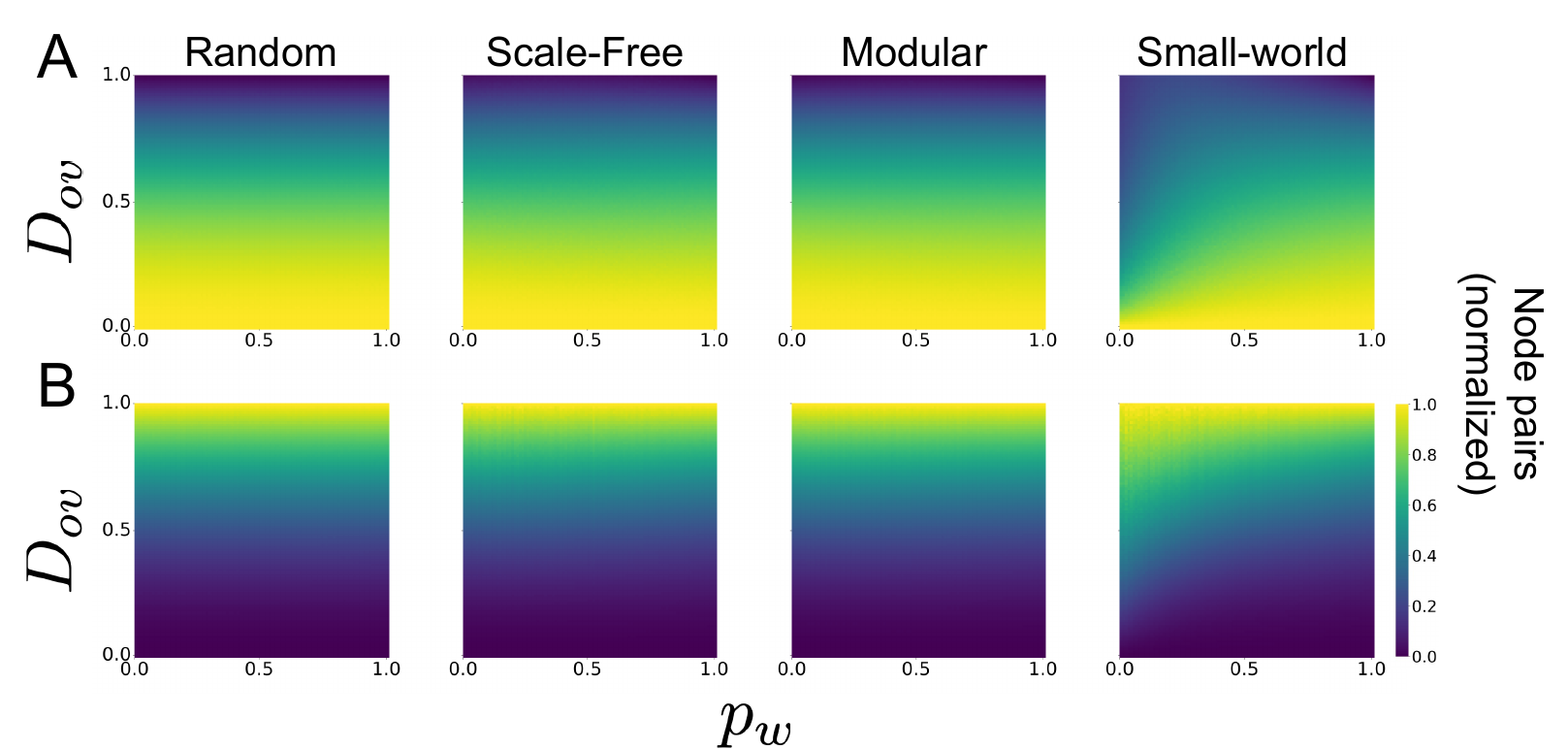}
    \caption{\textbf{Normalized distributions of unique and redundant node pairs.}
    The values shown in Fig.~\ref{s:s4} were normalized separately for each background network and each value of $p_w$ by the maximum observed across the full range of $D_{ov}$. 
    \textbf{(A)} Normalized unique node-pair counts and \textbf{(B)} normalized redundant node-pair
    counts. 
    This normalization highlights the opposing dependence of the two structural effects on overlap independently of differences in their absolute magnitudes.}
    \label{s:s5}
\end{figure}

\begin{figure}
\centering
\includegraphics[width=\textwidth]{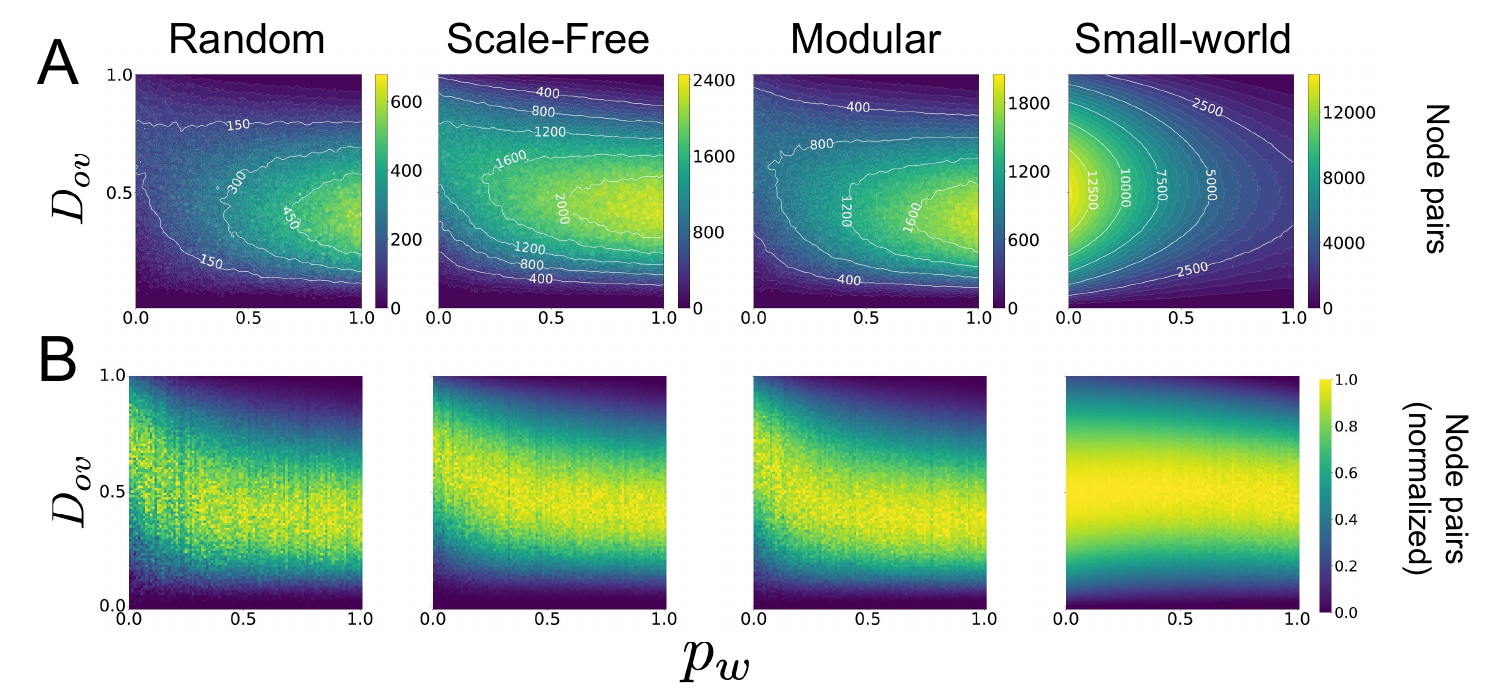}
    \caption{\textbf{Internal synergy after the integration of five individual knowledge networks.}
    \textbf{(A)} Absolute number of synergistic node pairs as a function of the within-individual wiring probability $p_w$ and overlap degree $D_{ov}$ for random, scale-free, modular, and small-world background networks.
    A node pair was classified as synergistic when integration produced a shorter path than was available in any individual network.
    \textbf{(B)} Corresponding values normalized, for each background network and each $p_w$, by the maximum across $D_{ov}$. 
    Separate color scales are used in panel A because the absolute number of synergistic pairs differs substantially among background-network structures.}
    \label{s:s6}
\end{figure}

\begin{figure}
\centering
\includegraphics[width=\textwidth]{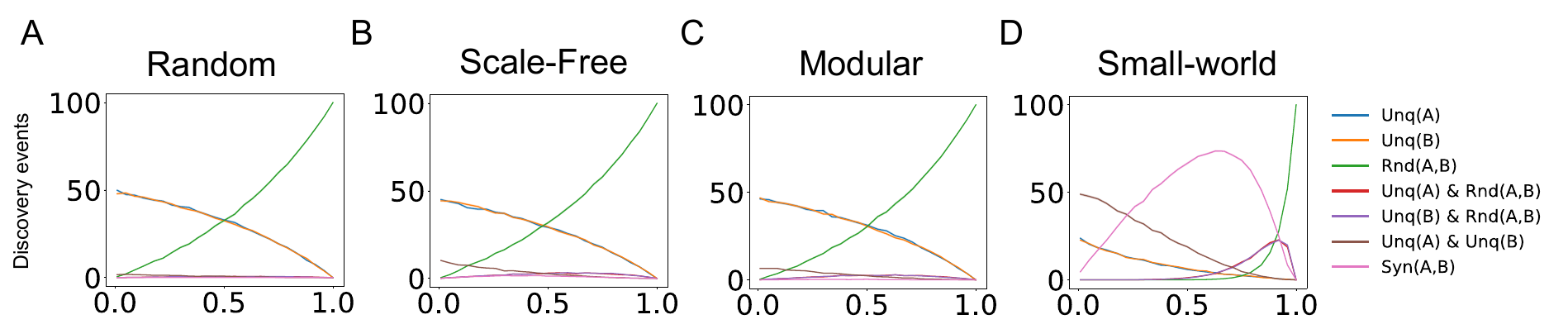}
    \caption{\textbf{Classification of external first-exit discovery events for two integrated individual networks.}
    Each discovered external node was classified according to its connections to $Unq(A)$, $Unq(B)$, and $Rnd(A,B)$. 
    Curves show the
    mean number of discovery events in the following classes:
    connection only to $Unq(A)$; connection only to $Unq(B)$; connection only to $Rnd(A,B)$; simultaneous connection to $Unq(A)$ and $Rnd(A,B)$; simultaneous connection to $Unq(B)$ and $Rnd(A,B)$; simultaneous connection to $Unq(A)$ and $Unq(B)$; and strict external synergy, $Syn(A,B)$, defined as simultaneous
    connection to all three regions. 
    Results are plotted against the overlap degree $D_{ov}$ for random, scale-free, modular, and small-world background networks.}
    \label{s:s7}
\end{figure}

\begin{figure}
\centering
\includegraphics[width=\textwidth]{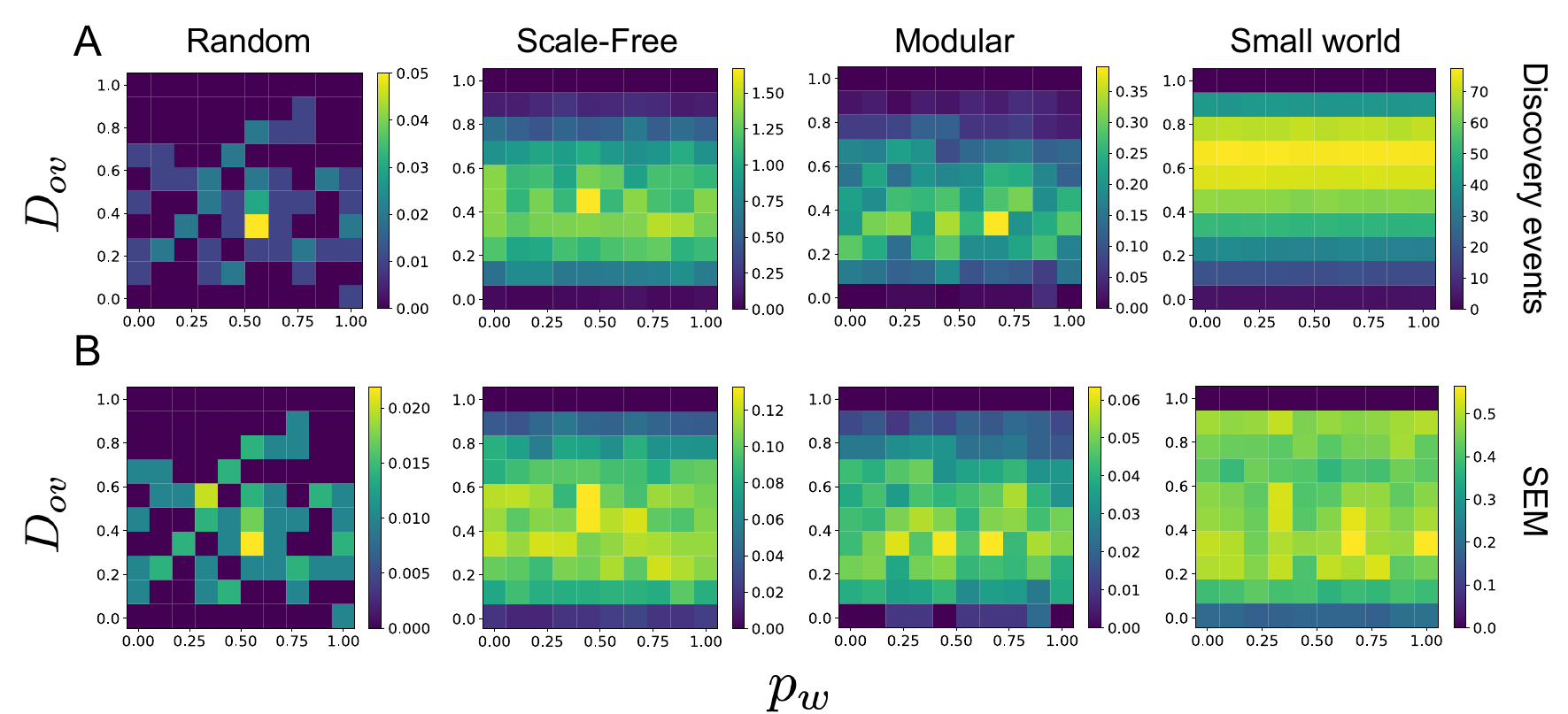}
    \caption{\textbf{Effects of within-individual connectivity and knowledge overlap on strict external-synergy discovery.}
    Heatmaps show \textbf{(A)} the mean number of strict external-synergy discovery events and \textbf{(B)} the corresponding SEM as functions of the within-individual wiring probability $p_w$ and overlap degree $D_{ov}$ for random, scale-free, modular, and small-world background networks.
    For each parameter combination, 100 external-node discoveries were generated by random walks in each of 100 repeats.
    An external node was classified as a strict synergy discovery when it was connected to the common region and to all individual-specific regions.
    Color scales are shown separately for each network topology.}
    \label{s:s8}
\end{figure}

\begin{figure}
\centering
\includegraphics[width=\textwidth]{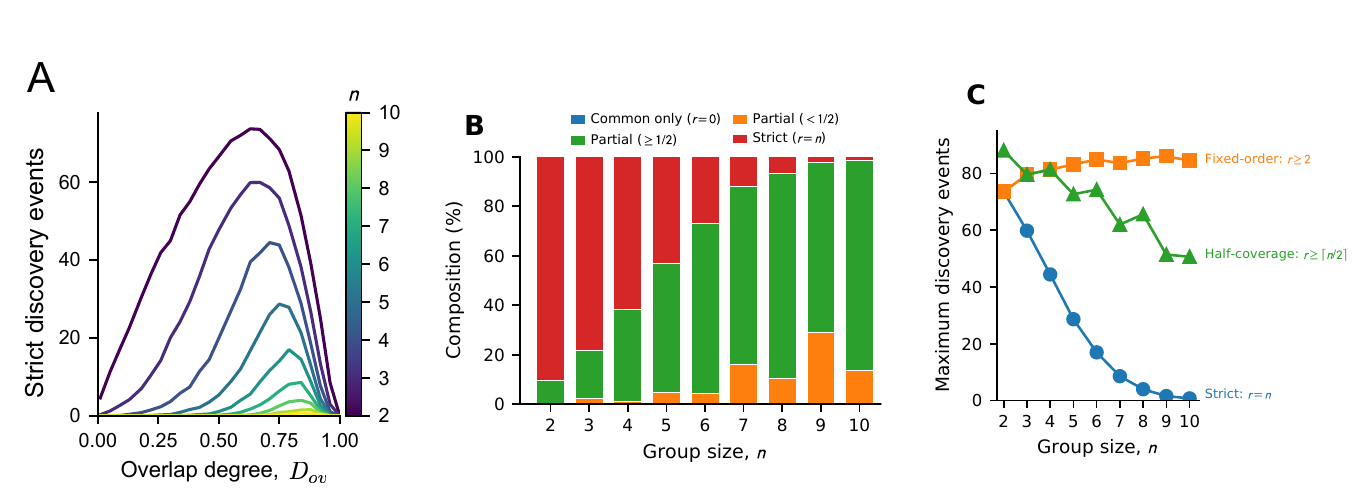}
    \caption{\textbf{Group-size dependence of strict and relaxed external-synergy criteria in small-world networks.}
    \textbf{(A)} Mean number of strict external-synergy discovery events as a
    function of overlap degree $D_{ov}$ for group sizes $n=2,\ldots,10$.
    For a group of $n$ individuals, the strict criterion requires a discovered
    external node to connect to the common region and to every
    individual-specific region ($b_R=1$ and $r=n$), where $r$ denotes the
    number of individual-specific regions connected to the external node.
    \textbf{(B)} Composition of discovery events connected to the common region,
    evaluated at the value of $D_{ov}$ that maximizes the strict criterion for
    each group size. Events are divided into common only ($r=0$), partial
    connections involving fewer than half of the individual-specific regions
    [$1\leq r<\lceil n/2\rceil$], partial connections involving at least half
    but not all regions [$\lceil n/2\rceil\leq r<n$], and strict connections
    ($r=n$). Bars are normalized to $100\%$. For $n=2$, the classes
    $r=0$, $r=1$, and $r=2$ correspond respectively to the redundant
    ($Rnd$), unique ($Unq$), and synergistic ($Syn$) categories used in the
    two-individual analysis. The common-only fraction is nonzero but very
    small and therefore appears as a thin band.
    \textbf{(C)} Maximum number of discovery events over $D_{ov}$ under three
    criteria: strict ($b_R=1$, $r=n$), fixed-order ($b_R=1$, $r\geq2$),
    and half-coverage [$b_R=1$, $r\geq\lceil n/2\rceil$].
    The maximum is evaluated independently over $D_{ov}$ for each criterion
    and group size.}
    \label{s:s9}
\end{figure}

\begin{figure}
    \centering
    \includegraphics[width=1.0\linewidth]{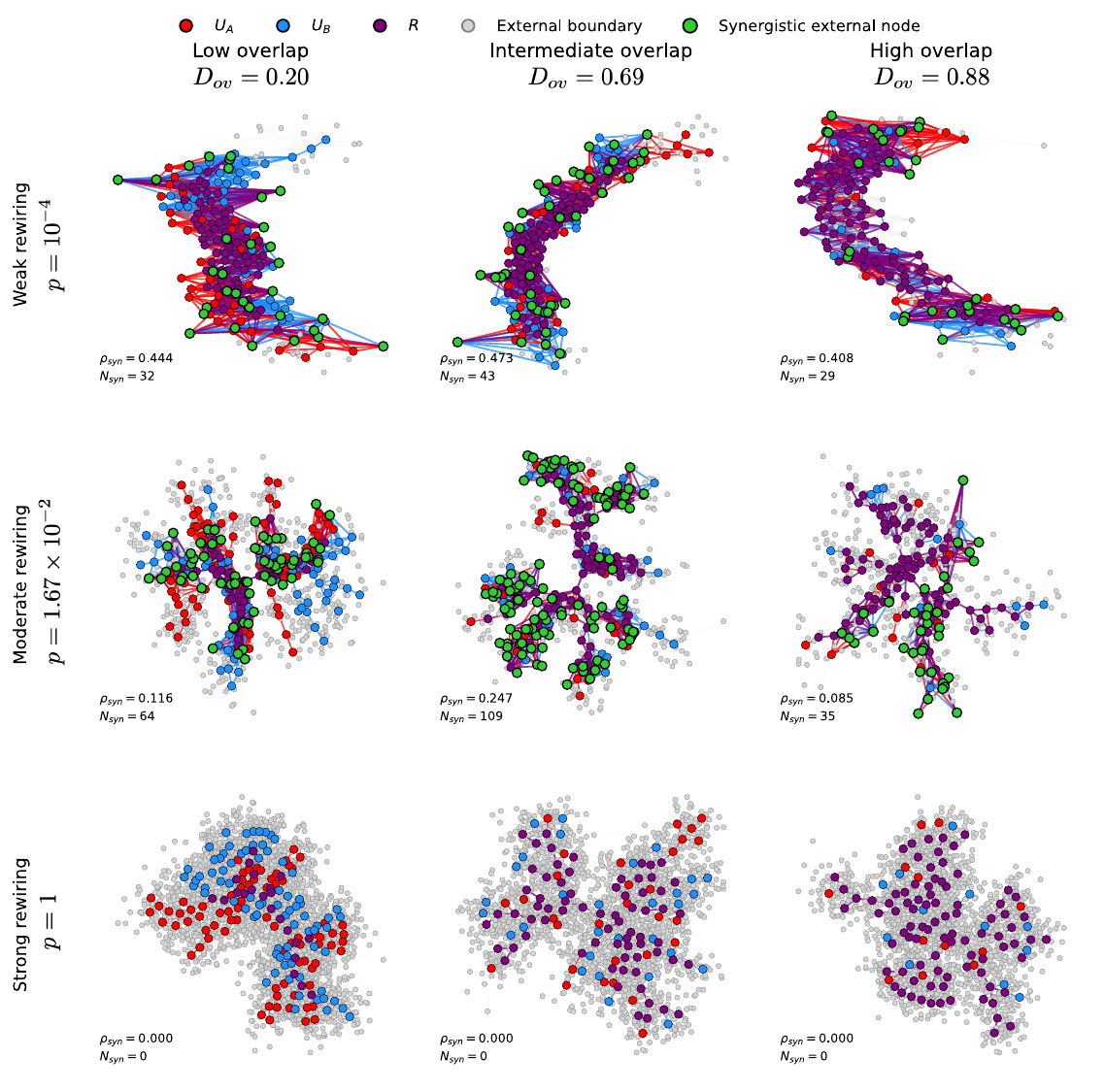}
    \caption{\textbf{Representative boundary organization across knowledge
    overlap and network rewiring.}
    Representative network realizations illustrate how the local boundary of
    the integrated knowledge network changes with overlap degree $D_{ov}$ and
    the Watts--Strogatz rewiring probability $p$.
    Columns show low ($D_{ov}=0.20$), intermediate ($D_{ov}=0.69$), and
    high ($D_{ov}=0.88$) overlap, and rows show weak
    ($p=10^{-4}$), moderate ($p=1.67\times10^{-2}$), and strong
    ($p=1$) rewiring.
    Red, blue, and purple indicate the A-specific region $U_A$, the
    B-specific region $U_B$, and the shared region $R$, respectively;
    internal knowledge relations follow the same color scheme.
    Gray nodes denote unknown external nodes adjacent to the integrated
    network, and green nodes denote structurally synergistic external nodes,
    defined as nodes having at least one edge to each of $U_A$, $U_B$, and
    $R$.
    Labels report the boundary concentration
    $\rho_{\mathrm{syn}}$ and the absolute number of
    structurally synergistic external nodes $N_{\mathrm{syn}}$ for each
    realization.
    All panels use $N_s=100$ and $p_w=0.5$.
    Each snapshot was selected from repeated realizations to have boundary
    concentration near the median for the corresponding parameter condition.
    The intermediate-overlap column ($D_{ov}=0.69$) is reproduced in
    Fig.~\ref{fig:fig4}D to isolate the effect of rewiring at fixed overlap.
    Node positions are used only for visualization and do not represent
    metric distances in the knowledge space.
    Quantitative changes in $N_{\mathrm{syn}}$,
    $\rho_{\mathrm{syn}}$, and realized synergistic discovery are
    reported in Fig.~\ref{fig:fig4}.}
    \label{s:s10}
\end{figure}

\begin{figure}
    \centering
    \includegraphics[width=1.0\linewidth]{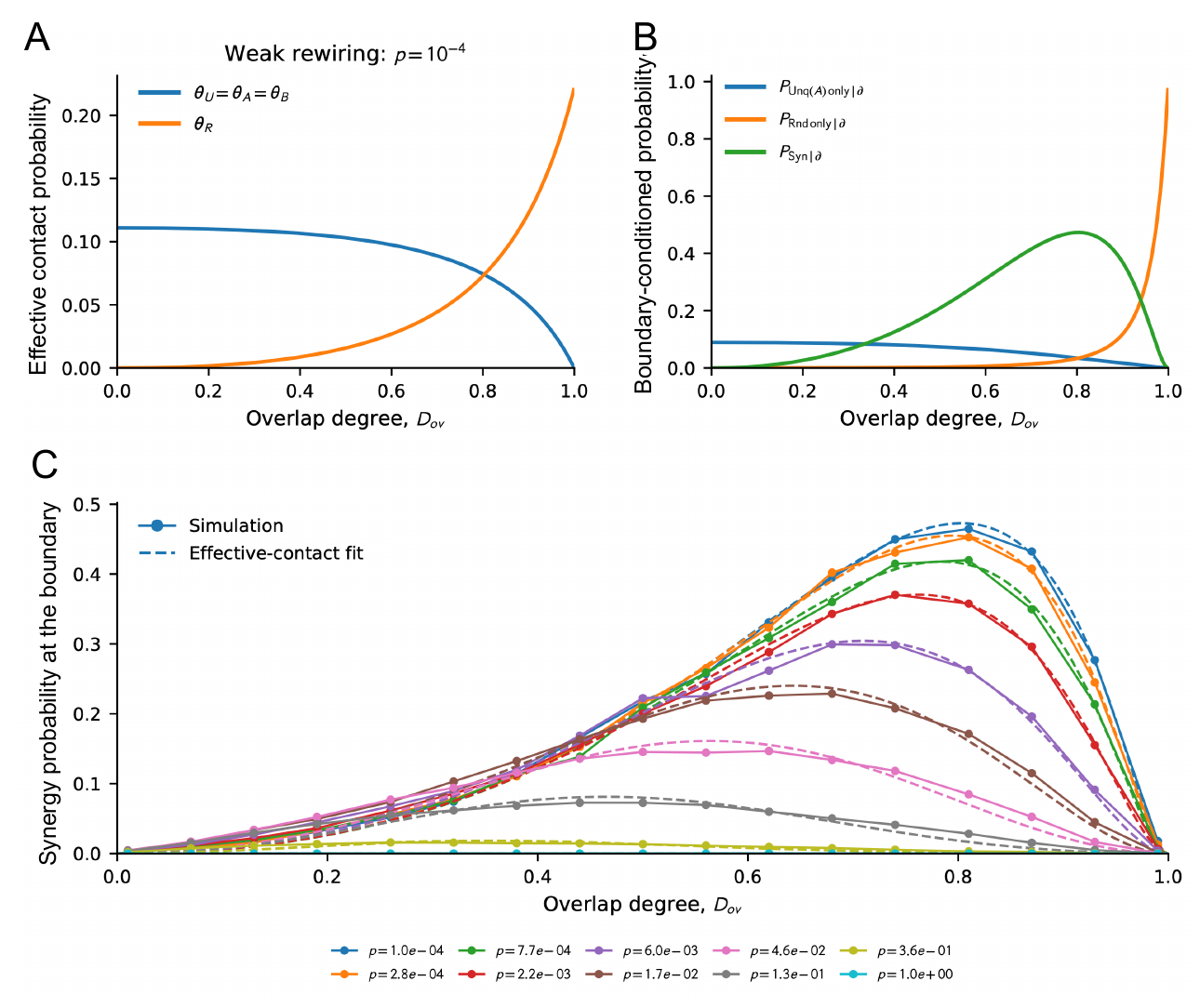}
    \caption{\textbf{Phenomenological effective-contact approximation of
    external synergy across overlap and rewiring.}
    \textbf{(A)} Effective per-contact probabilities with the
    individual-specific regions,
    $\theta_U=\theta_A=\theta_B$, and the shared region, $\theta_R$, for the
    weakly rewired condition $p=10^{-4}$.
    Increasing $D_{ov}$ shifts effective exposure from individual-specific
    toward shared knowledge.
    \textbf{(B)} Boundary-conditioned probabilities of connection only to
    the A-specific region, only to the shared region, and simultaneously to
    both individual-specific regions and the shared region under the
    effective-contact approximation. The strict three-region condition
    produces an intermediate-overlap maximum.
    \textbf{(C)} Direct comparison between simulated boundary concentration
    $\rho_{\mathrm{syn}}$ and the phenomenological approximation
    $P_{\mathrm{Syn}\mid\partial}$ across rewiring probabilities.
    Solid curves with markers show the simulation results corresponding to
    Fig.~\ref{fig:fig4}C, whereas dashed curves show the effective-contact reconstruction
    obtained by fitting $\phi_p$ and $\eta_p$ separately for each $p$.
    The approximation captures the decrease in peak magnitude and the
    approximate shift of the peak toward lower $D_{ov}$ as rewiring increases.
    Because the parameters are fitted directly to the boundary-concentration
    curves, the dashed lines represent phenomenological reconstructions rather
    than independent theoretical predictions.}
    \label{s:s11}
\end{figure}

\end{document}